\documentclass[aps,prl,preprint,groupedaddress]{revtex4-1}
\usepackage{color}
\draft % marks overfull lines with a black rule on the right
\usepackage{graphicx}
\usepackage{caption}
\usepackage{float}
\begin{document}

%\preprint{}

\title{Modeling of heterojunction photovoltaic cells based on ZnO nanowires array and earth-abundant cuprous oxide absorbers}


\author{Qilin Gu}

\email[]{gump423@gmail.com}
\homepage[]{https://sites.google.com/site/qlgu423/}
%\thanks{}
\affiliation{Department of Electrical and Computer Engineering, The Ohio State University, Columbus, OH 43210, USA}


\date{November 26, 2010}

\begin{abstract}
As a potential solution for low-cost efficient solar cells, radial junctions consisting of ZnO nanowires arrays embedded in $\mathrm{Cu_2O}$ thin films have been theoretically modeled. Calculations have been performed to explore the geometric dependence of performance of such wire-based solar cells. By properly setting material properties and cell dimensions, a reasonable power conversion efficiency of 19.7\% can be expected in a material with 2 $\mu$m minority carrier diffusion length. The detrimental effects of bulk, interface and contact-related states on solar cell performance have also been studied, from which the efficiencies between $\scriptsize{\sim}$22\% and $\scriptsize{\sim}$12\% for a series of materials, ranging from optimal to seriously poor-quality, are extracted. The findings suggest that rational device design plays a crucial role in implementing efficient $\mathrm{Cu_2O}$/ZnO wire radial junction solar cells.
\end{abstract}

%\keywords{}

\maketitle

%\section{}
Among various novel photovoltaic schemes that try to address cost and efficiency issues, solar cells based on semiconductor nanowires (NWs) and nanopillars (NPLs) are relatively new attempts. It has been demonstrated, both theoretically and experimentally, that NW/NPL array 3D structures can simultaneously enhance the optical absorption and photo-generated carrier collection by light trapping and decoupling of light absorption and photo-carrier separation respectively.\textcolor{blue}{$^{1-4}$} At the early stage, most remarkable efforts in this field so far have been devoted to the studies of silicon NW/NPL based photovoltaic devices.\textcolor{blue}{$^{3-8}$} Theoretically, the advantages of NW/NPL array structures become more significant when the materials with poor minority carrier diffusion lengths are used.\textcolor{blue}{$^1$} Therefore, nanowires and nanopillars of low-cost thin film materials have become increasingly important. 3D CdS/CdTe NPL-array solar cells with efficiency of $\scriptsize{\sim}$6\% have been successfully fabricated,\textcolor{blue}{$^9$} and the corresponding theoretical study indicated optimal devices with $>$20\% efficiency could be achieved.\textcolor{blue}{$^{10}$}

With its high absorption coefficient and good minority carrier diffusion length, earth-abundant and non-toxic $\mathrm{Cu_2O}$ has always been considered a promising material for low-cost photovoltaic applications. The heterojunction based on \emph{p}-$\mathrm{Cu_2O}$ thin film and \emph{n}-ZnO NW/NPL arrays has recently been attractive due to the ease of preparation and its potential for reasonable efficiency. Despite the research efforts on experimental demonstrations of $\mathrm{Cu_2O}$/ZnO-NWs,\textcolor{blue}{$^{11, 12}$} the understanding of constraints that limit the cell efficiency has yet been well-developed. In this letter, the effects of material properties and interface states on the performance of $\mathrm{Cu_2O}$/ZnO-NWs solar cells are explored through physical device modeling to extend our knowledge to thin film radial heterojunctions. 

As shown in Fig. 1(a) and (b), the radial junction NWs solar cell structure employed in this work consists of an array of highly doped \emph{n}-type ZnO nanowires array embedded in the surrounding \emph{p}-type $\mathrm{Cu_2O}$ thin film absorber layer. A transparent conducting oxide (TCO) and Al metal layer are considered as top and bottom contact respectively, and a 0.2 $\mu$m insulator is used to separate the bottom contact from the absorber. The effective thickness of the wire-based solar cell \emph{L}, which represented the length of the overlapping part between $\mathrm{Cu_2O}$ and ZnO, and the total radius of the unit cell (as shown in Figure 1(b)) \emph{R} are dimensional parameters that need to be optimized in this study. 

The physical device modeling was implemented by numerically solving the complete set of basic semiconductor equations of a \emph{p-n} heterojunction (band structure shown in Figure 1(d)) including Poisson's equation, continuity equations and drift-diffusion transport equations simultaneously and self-consistently in a two-dimensional (2D) scenario using finite element methods. The major difference here from conventional device modeling is that the solutions to carrier transport equations were obtained by extending the analysis of planar solar cell to a cylindrical geometry. In our modeling, the light with AM1.5G spectrum was assumed to be normally incident on the top surface of the TCO layer where the absorption was considered negligible, without considering any reflection losses. Besides, due to the very high bandgap of ZnO nanowires, it was also assumed that light absorption and photo-generation of carriers only occurred in $\mathrm{Cu_2O}$ layer whose absorption coefficient and corresponding optical depth were plotted in Fig. 1(c).\textcolor{blue}{$^{13}$} It is worth noting that $\scriptsize{\sim}$100 $\mu$m thickness is needed to absorb most of the light with photon energy near the $\mathrm{Cu_2O}$ band edge. Recombination was assumed to be solely due to the Shockley-Read-Hall (SRH) recombination from a single mid-gap state, with Auger and surface recombination processes neglected. In addition, the minority carrier transport process was assumed to be purely along the radial direction for simplicity. This 1-D approximation is valid in a collection-limited material with optical depth greatly exceeding the minority carrier diffusion length $L_n$, which was true for our $\mathrm{Cu_2O}$ case. The preliminary results were obtained based on the above assumptions by setting the primary material quality parameter $L_n$=2 $\mu$m, a moderately large minority carrier diffusion length for cuprous oxide , in the doped absorber with hole concentration of 1$\times$$\mathrm{10^{17}}$ $cm^{-3}$, a reasonable doping level for $\mathrm{Cu_2O}$ thin films. The ZnO NWs cores were designed to be relatively heavily doped with $N_d$=5$\times$$\mathrm{10^{18}}$ $cm^{-3}$ so that parasitic resistance could be minimized. The resultant moderate depletion widths can also help to avoid interactions of depletion regions between neighboring unit cells. 

The power conversion efficiency (PCE) as a function of geometry is mapped in Figure 2(a), where the effects of the length \emph{L} and radius \emph{R} of the NWs unit cells can be evaluated. The peak cell efficiency $\eta$ of 19.7\% occurs for NWs solar cells with cell thickness \emph{L} of approximately optical thickness $\scriptsize{\sim}$100 $\mu$m and the total cell radius \emph{R} of 1 $\mu$m - half of the minority carrier diffusion length. The corresponding illuminated \emph{I-V} curve is also plotted in Fig. 2(b) to clearly show the short-circuit current $J_{SC}$ and open-circuit voltage $V_{OC}$ for the maximum $\eta$. Comparisons are made for \emph{I-V} curves of NWs cells with different radius \emph{R} at the same cell thickness of \emph{L}=100 $\mu$m, from which it is found that the efficiency drops to 7.9\% when \emph{R} increases to 10 $\mu$m. By noticing the 16.3\% efficiency of planar junction solar cell consisting of ZnO and $\mathrm{Cu_2O}$ thin films with non-optimal layer thicknesses (0.2 $\mu$m ZnO and 5 $\mu$m $\mathrm{Cu_2O}$), it's worth mentioning that lack of optimal designs of wire-based solar cells could potentially fail to achieve superior performance to the their planar counterparts. From Figure 2(c), we can tell that short-circuit current $J_{SC}$ increases, as expected, with increasing cell thickness \emph{L} and consequently increasing photo-generated carriers, saturating when \emph{L} reaches the optical thickness. Also, $J_{SC}$ seems essentially independent of cell radius \emph{R} when \emph{R} is within the range of $L_n$ and starts to drop sharply if \emph{R} $>$ $L_n$, which can be reasonably attributed to the significant recombination-induced carrier loss caused by small diffusion length relative to the long minority carrier traveling path. On the other hand, $V_{OC}$ decreases with increasing \emph{L} and increases with increasing \emph{R}. The open-circuit voltage is lost due to the geometric increase of \emph{p-n} junction interface area and the subsequent decrease in the ratio of photocurrent to dark current, when cell thickness increases. In addition, the enhancement of cell radius can yield higher excess carrier concentration, thereby improving $V_{OC}$. Therefore, the aforesaid $J_{SC}$ and $V_{OC}$ trend, along with fill factor which roughly follows voltage behavior, determine the overall efficiency depending on different cell dimensions. 

To explore the effect of bulk defect states on NWs solar cell performance, the similar calculations were carried out for different minority carrier diffusion length Ln which can be used to representatively describe the density of bulk states. Figure 3(a) shows the cell radius dependence of efficiencies for ZnO/$\mathrm{Cu_2O}$ coaxial NWs solar cells with $L_n$ ranging from 0.2-5 $\mu$m and cell thickness of 100 $\mu$m for sufficient light absorption. Clearly the peak efficiency goes down when $L_n$ is decreased due to the overall enhancement of carrier recombination process before collection. Differences can also be observed for the geometric dependence of efficiency for different $L_n$: The peak efficiency shifts to smaller cell radius values with decreasing $L_n$, which is reasonable since reduced radial distances are more favored for low $L_n$ cases to avoid serious recombination. Note that up to 22\% power conversion efficiency can be expected in a material with minority diffusion length $L_n$=5 $\mu$m, and even a poor-quality absorber layer with much lower $L_n$=0.2 $\mu$m can still lead to a moderate peak efficiency of $\scriptsize{\sim}$13\%, suggesting a promising future of oxide-based wire solar cells. The corresponding $J_{SC}$ and $V_{OC}$ behaviors when varying minority diffusion length are also plotted as functions of $L_n$ and \emph{R} in Fig. 3(b). The slight dependence of $J_{SC}$ on $L_n$ for small fixed cell radius, which is contrast to planar junction case, is expected for radial cell geometry since excess minority carriers can be efficiently collected when almost all of them are generated within diffusion length from the core/shell interface. While $V_{OC}$ degrades more rapidly with poor material quality, resulting in major efficiency losses. On the other hand, to evaluate dependence of cell performance on \emph{p-n} heterojunction interface quality whose importance becomes significant since interfacial area increases in wire-based radial cells, performance parameters $\eta$, $J_{SC}$, $V_{OC}$ and \emph{FF} are plotted as a function of 2-D sheet interface states density $N_{I, S}$ in Fig. 3(c). For this and further calculations, a cell with $L_n$=2 $\mu$m, \emph{L}=100 $\mu$m, \emph{R}=1 $\mu$m is used as the model configuration. The increase of interface states density doesn't seriously alter the efficiency when $N_{I, S}$ is relatively low, while $\eta$ drops dramatically after density exceeds $10^{10}$ $cm^{-2}$, with $\eta$~14.2\% at $N_{I,S}$=$10^{13}$ $cm^{-2}$. With $J_{SC}$ remaining nearly constant, the efficiency loss is found to primarily result from the decrease of $V_{OC}$ due to the increased dark current caused by enhanced interfacial recombination. 

Finally, we focus the interface states at TCO/\emph{p}-$\mathrm{Cu_2O}$ top contact where double-fold impact on device performance is considered. Practically, as shown in the upper inset of FIG. 4(a), high densities of interface states at the contacts between TCO and \emph{p}-type semiconductors can cause Fermi level pinning and consequently form substantial Schottky potential barriers, which could be a drawback for efficient carrier collections.\textcolor{blue}{$^{14}$} This effect is studied by modeling an opposite diode connected to the primary \emph{p-n} junction. Fig. 4(a) shows the dependence of conversion efficiency on barrier height $\Phi_p$. Under low barrier height conditions, $\eta$ only shows mild reduction from $\scriptsize{\sim}$20\% to $\scriptsize{\sim}$19\% when $\Phi_p$$<$0.4 eV, but it decreases drastically to 12\% when $\Phi_p$ increases from 0.6 eV to higher values. From the lower inset of Figure 4(a), It is evident that significant $V_{OC}$ loss, rather than decrease of $J_{SC}$, is predominantly responsible for the efficiency degradation. Besides, fill factor (\emph{FF}, not shown here) drops significantly with increasing barrier height $\Phi_p$, which also contributes to the energy loss induced by the unexpected rectifying barrier, in agreement with previous studies of barrier height in solar cell contacts.\textcolor{blue}{$^{10, 15}$} The other negative effect is decreasing cell efficiencies through the recombination at the interface states of top contacts. The light \emph{I-V} curves for radial wire cells with increasing sheet density of top contact interface states $N_{C, S}$ from $10^9$-$10^{14}$ $cm^{-2}$, as plotted in Fig. 4(b), reveal an obvious $J_{SC}$ reduction due to the enhanced photo-carrier recombination at top contacts induced by increased $N_{C, S}$, resulting in the strong dependence of efficiency on interface states density at top contact and serious $\Phi_p$ degradation to $\scriptsize{\sim}$13\% when $N_{C, S}$=$10^{14}$ $cm^{-2}$ (inset of Fig. 4(b)).

In summary, the predicted performance of radial junction based on $\mathrm{Cu_2O}$/ZnO NWs is investigated by modeling the dependence of cell parameters, including efficiency, short-circuit current and open-circuit voltage, on materials properties and geometric dimensions. The results reveal that optimization of material quality and device geometry can lead to an attainable power conversion efficiency of $\eta$ $\scriptsize{\sim}$22\%, indicating the role of device modeling and design. Moreover, the studies also show the degradation of cell performance with high interface states densities at $\mathrm{Cu_2O}$/ZnO heterojunction and TCO/$\mathrm{Cu_2O}$ top contact, addressing the importance of controlling the quality of critical interfaces.
\vfill
\newpage
\noindent
\underline{\textbf{Reference}}\\
$^1$B. M. Kayes, H. A. Atwater, and N. S. Lewis, J. Appl. Phys. \textbf{97}, 114302 (2005).\\
$^2$L. Hu and G. Chen, Nano Lett. \textbf{7}, 3249 (2007).\\
$^3$E. Garnett and P. Yang, Nano Lett. \textbf{10}, 1082 (2010).\\
$^4$M. D. Kelzenberg, S. W. Boettcher, J. A. Petykiewicz, D. B. Turner-Evans, M. C. Putnam, E. L. Warren, J. M. Spurgeon, R. M. Briggs, N. S. Lewis and H. A. Atwater, Nature Mater. \textbf{9}, 239 (2010). \\
$^5$L. Tsakalakos, J. Balch, J. Fronheiser, B. A. Korevaar, O. Sulima, and J. Rand, Appl. Phys. Lett. \textbf{91}, 233117 (2007).\\
$^6$K. Peng, X. Wang, and S.-T. Lee, Appl. Phys. Lett. \textbf{92}, 163103 (2008).\\
$^7$O. Gunawan and S. Guha, Sol. Energy Mater. Sol. Cells \textbf{93}, 1388 (2009).\\
$^8$M. Putnam, S. Boettcher, M. D. Kelzenberg, D. Turner-Evans, J. Spurgeon, E. Warren, R. Briggs, N. Lewis and H. Atwater, Energy Environ. Sci. \textbf{3}, 1037 (2010).\\
$^9$Z. Fan, H. Razavi, J. Do, A. Moriwaki, O. Ergen, Y.-L. Chueh, P. W. Leu, J. C. Ho, T. Takahashi, L. A. Reichertz, S. Neale, K. Yu, M. Wu, J. W. Ager, and A. Javey, Nature Mater. \textbf{8}, 648 (2009).\\
$^{10}$R. Kapadia, Z. Fan, and A. Javey, Appl. Phys. Lett. \textbf{96}, 103116 (2010).\\
$^{11}$J. Cui and U. J. Gibson, J. Phys. Chem. C \textbf{114}, 6408(2010).\\
$^{12}$K. P. Musselman, A. Wisnet, D. C. Iza, H. C. Hesse, C. Scheu, J. L. MacManus Driscoll, and L. Schmidt-Mende, Adv. Mater. \textbf{22}, E254 (2010).\\
$^{13}$P. W. Baumeister, Phys. Rev. \textbf{121}, 359 (1961).\\
$^{14}$F. Sanchez Sinencio and R. Williams, J. Appl. Phys. \textbf{54}, 2757 (1983).\\
$^{15}$S. H. Demtsu, J. R. Sites, Thin Solid Films \textbf{510}, 320 (2006).\\

\vfill
\newpage
\begin{figure}[H]
\centering
\includegraphics[width=0.8\textwidth]{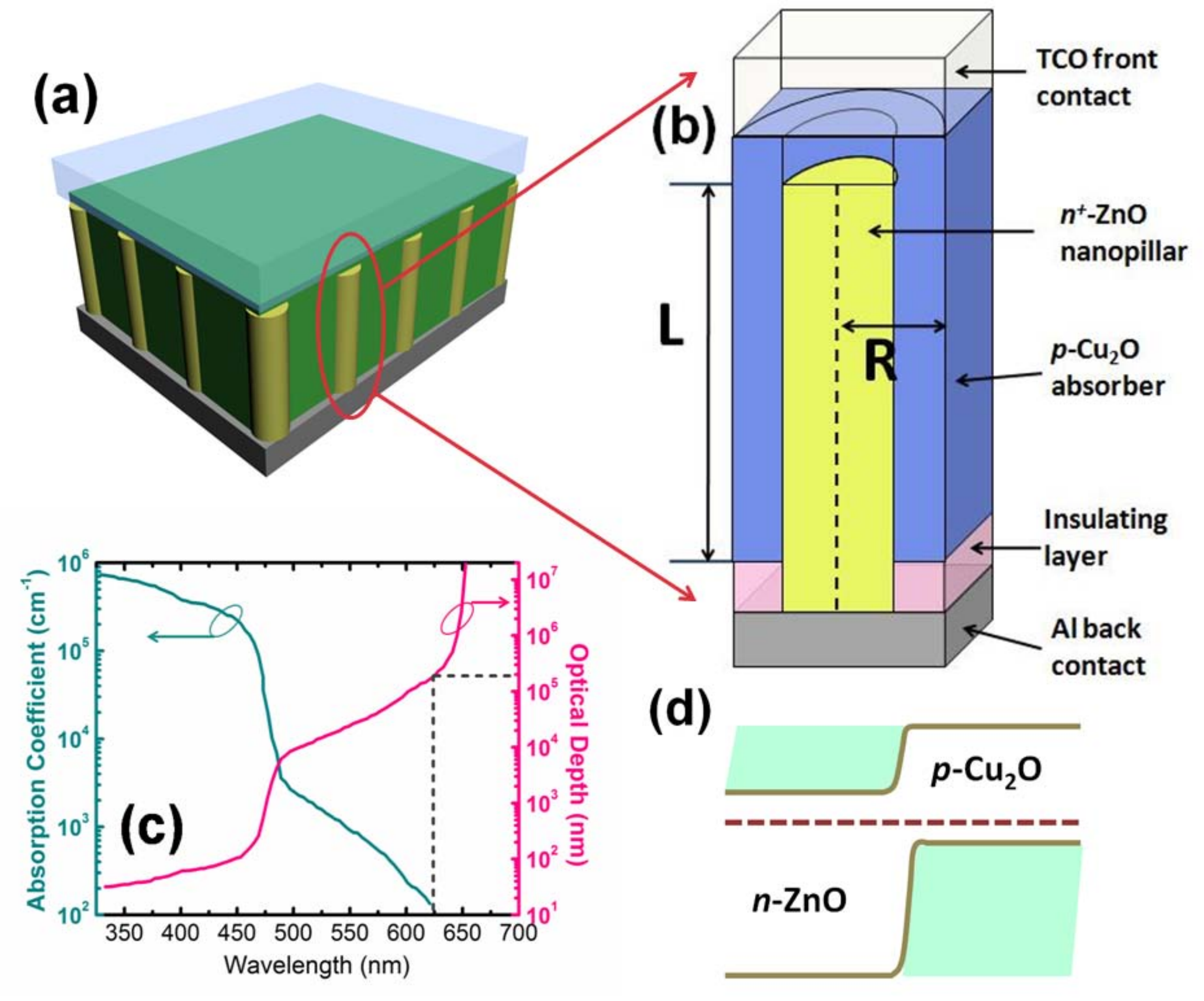}
\caption{\label{FIG.1} Schematic wires-based heterojunction structure: (a) a prospective view of a cell consisting of $\mathrm{Cu_2O}$ thin film and an array of ZnO nanowires with (b) detailed configuration of a unit cell. (c) Absorption properties of $\mathrm{Cu_2O}$ absorber. (d) Simplified band diagram of the heterojunction under study.}
\end{figure}
\vfill
\newpage

\begin{figure}[H]
\centering
\includegraphics[width=1\textwidth]{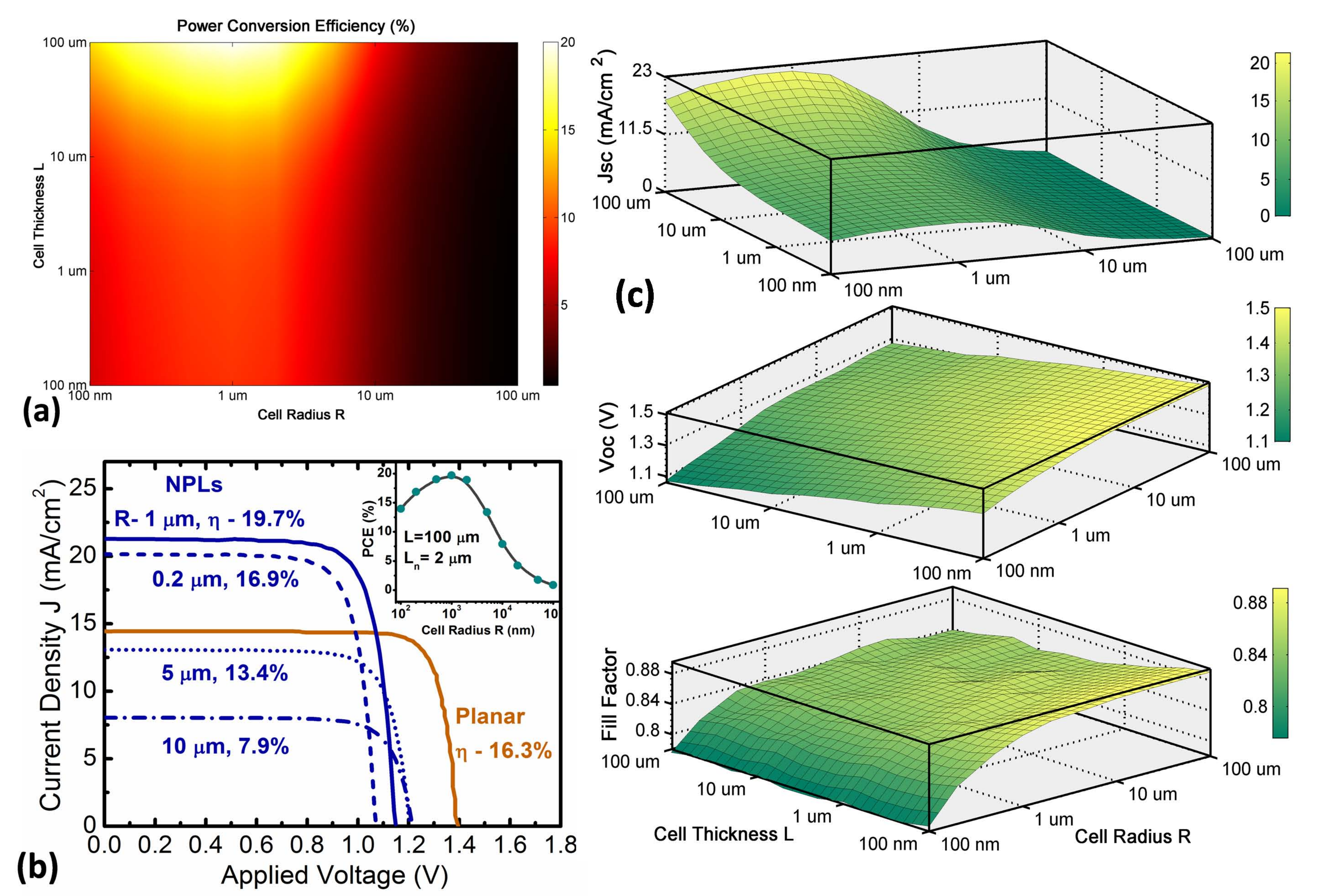}
\caption{\label{FIG.2} The geometric dependence of power conversion efficiency (PCE) (a), short-circuit current, open-circuit voltage and fill factor (c) of the studied $\mathrm{Cu_2O}$/ZnO wire solar cells. The light $I-V$ curves for wire cells with 100 $\mu$m thickness and radius of 0.2, 1, 5, 10 $\mu$m, along with the one for $\mathrm{Cu_2O}$/ZnO planar junction, were plotted in (b) whose inset show the cell radius dependence of cell PCE.}
\end{figure}
\vfill
\newpage
\begin{figure}[H]
\centering
\includegraphics[width=1\textwidth]{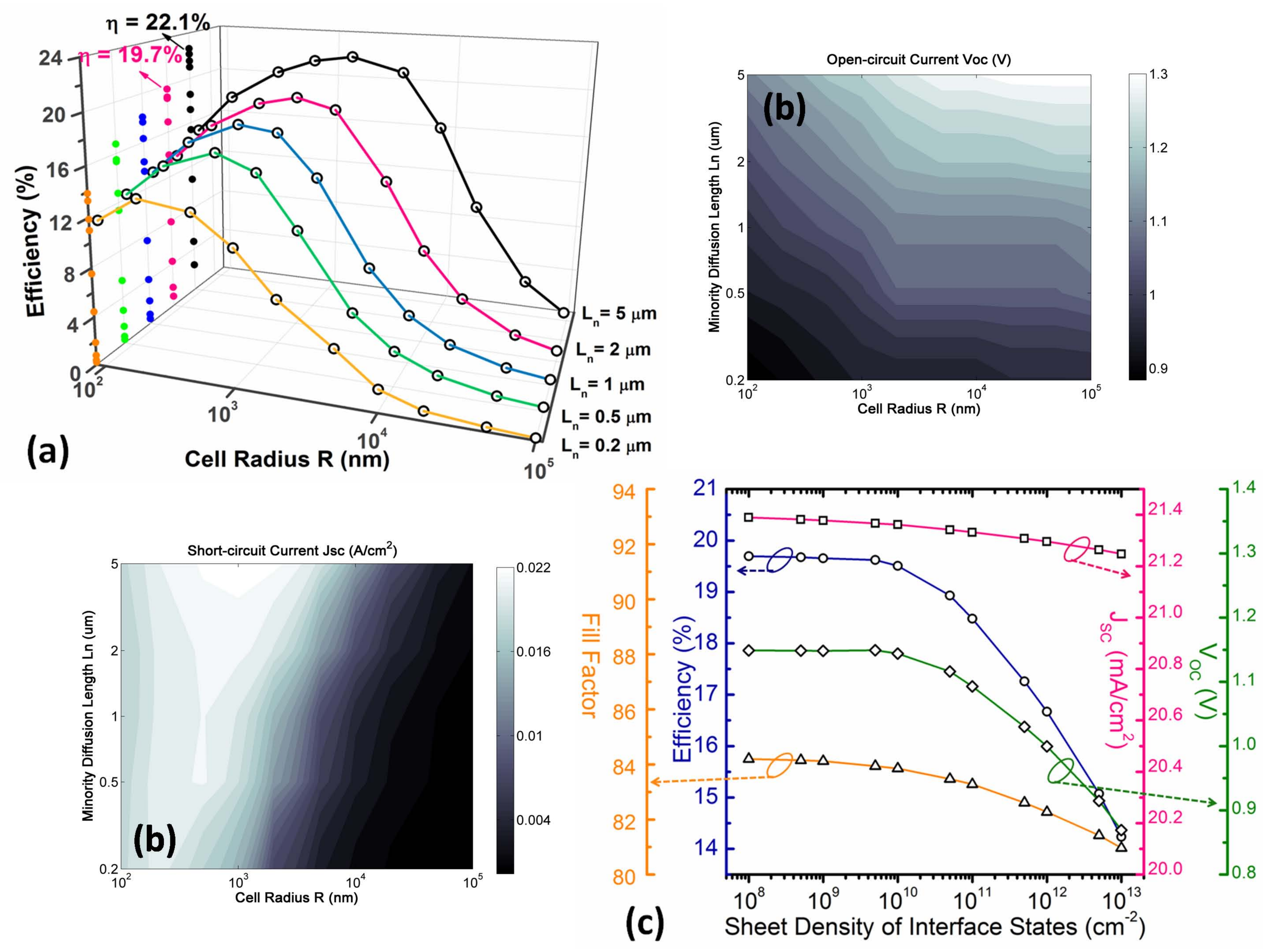}
\caption{\label{FIG.3}The effects of bulk defect states and interface states on wire cell performance: (a) efficiencies $\eta$ of cells with $L_n$ ranging from 0.2-5 $\mu$m were plotted as a function of $R$, the cell thickness is fixed to be 100 $\mu$m to achieve sufficient absorption. (b) Contour plots of short-circuit current $J_{SC}$ (lower-left) and open-circuit voltage $V_{OC}$ (upper-right) as functions of $L_n$ and $R$. (c) Plots of $\eta$, $J_{SC}$, $V_{OC}$ and fill factor (FF) as a function of sheet density of interface states $N_{I,S}$,}
\end{figure}
\vfill
\newpage

\begin{figure}[H]
\centering
\includegraphics[width=1\textwidth]{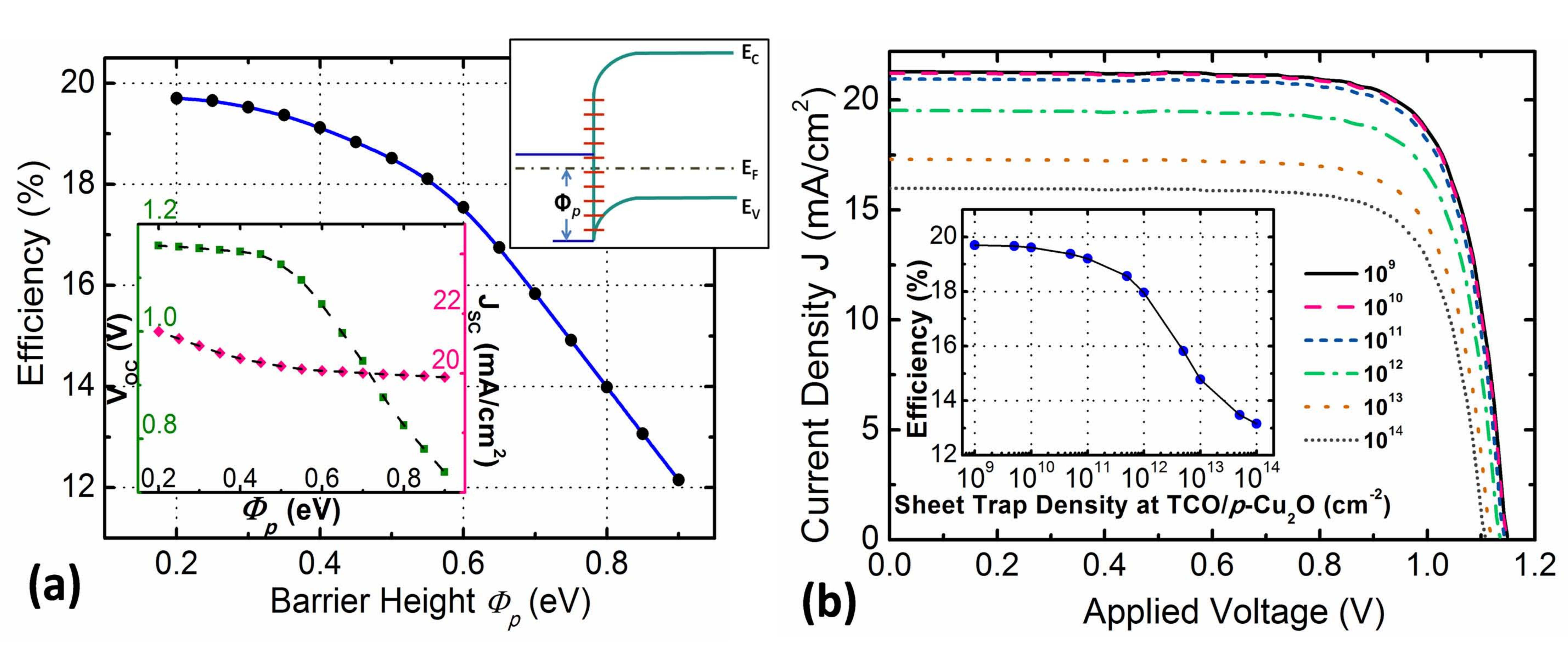}
\caption{\label{FIG.4} Cell efficiency as a function of barrier height $\Phi_p$ at the TCO/$p-\mathrm{Cu_2O}$ contact, the lower inset shows the barrier height dependences of $J_{SC}$ (pink diamonds) and $V_{OC}$ (green squares) and the upper inset shows the schematic diagram of band structures near the TCO/$Cu_2O$ interface. (b) Light $I-V$ curves for wire cells with sheet density of TCO/$\mathrm{Cu_2O}$ interface states $N_{C, S}$ ranging from $10^9$-$10^{14}$ $cm^{-2}$, which revealed $J_{SC}$ and $V_{OC}$ for each case. $N_{C, S}$ dependence of efficiency was plotted in the inset. $L$=100 $\mu$m, $R$=1 $\mu$m, $L_n$=2 $\mu$m and $N_{I, S}$=$10^8$ $cm^{-2}$ were used for both calculations.}
\end{figure}

\subsection{}
\subsubsection{}

\bibliography{}

\end{document}